\documentclass[12pt]{article}

\usepackage{mathptmx}
\usepackage{algorithm}
\usepackage[noend]{algpseudocode}
\usepackage{amsmath, amssymb, amsthm}
\usepackage{graphicx}
\usepackage[super,comma,sort&compress]{natbib}
\usepackage{booktabs}
\usepackage{array}
\usepackage{multirow}
\usepackage{mdframed}
\newcounter{algocount}

\usepackage{setspace}
\usepackage{geometry}
\usepackage[hidelinks]{hyperref}
\usepackage{xcolor}
\usepackage{enumitem}
\usepackage{titling}
\usepackage{float}
\usepackage[utf8]{inputenc}
\usepackage[normalem]{ulem}   
\usepackage{soul}

\usepackage{tikz}
\usetikzlibrary{positioning, arrows.meta, fit, backgrounds, calc}

\usepackage[section]{placeins}

\begin{document}
\setstretch{1}

\begin{center}

{\Large\bfseries
Privacy-preserving causal mediation analysis using distributed electronic health record networks
\par}

\vspace{1.2cm}

{\normalsize
Hyojung Jang\textsuperscript{1,\ensuremath{\dagger}},
Rotana Radwan\textsuperscript{2,\ensuremath{\dagger}},
Malcolm Risk\textsuperscript{3},
Yao Lee\textsuperscript{4},
Jiang Bian\textsuperscript{4},
Xu Shi\textsuperscript{3},\\
Serena Guo\textsuperscript{2,\ensuremath{\ddagger}},
and Lili Zhao\textsuperscript{1,\ensuremath{\ddagger},*}
\par}

\vspace{0.5cm}

{\small
\textsuperscript{1}Division of Biostatistics, Department of Preventive Medicine, Northwestern University, Chicago, IL, USA\\
\textsuperscript{2}College of Pharmacy, Purdue University, West Lafayette, IN, USA\\
\textsuperscript{3}Department of Biostatistics, University of Michigan, Ann Arbor, MI, USA\\
\textsuperscript{4}Regenstrief Institute, Indianapolis, Indiana, USA\\
\par}

\vspace{0.5cm}

{\footnotesize
\textsuperscript{\ensuremath{\dagger}}Rotana Radwan is a co-first author\\
\textsuperscript{\ensuremath{\ddagger}}Serena Guo is a co-Senior \\
\textsuperscript{*}Correspondence: Lili Zhao; \texttt{[zhaolili@northwestern.edu]}.
\par}

\end{center}

\vspace{0.8cm}

\bigskip

\begin{abstract}
Electronic health record (EHR) networks provide unprecedented opportunities to study treatment mechanisms at scale, but mediation analyses across institutions are often hindered by privacy and governance constraints that restrict sharing of patient-level data. We developed a privacy-preserving federated mediation framework that enables estimation of natural direct and indirect effects without exchanging individual-level records across participating sites. The proposed approach integrates renewable learning with counterfactual causal mediation analysis, allowing institutions to collaboratively investigate treatment mechanisms using only low-dimensional summary statistics. Both simulation studies and the real-world application demonstrated that the federated estimator closely reproduced pooled-data results while preserving patient privacy. We applied the method to 32,146 patients in the Indiana Network for Patient Care to evaluate the extent to which body mass index (BMI) mediates the effect of GLP-1 receptor agonist  on glycated hemoglobin (HbA1c) reduction. The BMI-mediated pathway accounted for only a small proportion of the overall treatment effect, suggesting that most glycemic improvement occurred through mechanisms other than weight loss.

\bigskip

\end{abstract}

\newpage
\setstretch{1.9}

\section{Introduction}
\label{sec:intro}

Understanding not only whether a treatment works, but how it works, is a central aim
of clinical and epidemiologic research. Causal mediation analysis provides a rigorous framework for characterizing the pathways through which a treatment affects an outcome by decomposing the total effect into a direct effect and an indirect effect operating through an intermediate variable (mediator) \citep{robins1992,pearl2001,vanderweele2015}. Such decompositions can reveal whether treatment benefits arise through changes in biomarkers, behaviors, or downstream physiological processes, thereby generating hypotheses about modifiable mechanisms and informing the development of more targeted  interventions\citep{vanderweele2016,steen2017}. The widespread adoption of real-world data
infrastructures, including electronic health records (EHRs) and administrative claims
\citep{sherman2016,concato2022}, has greatly expanded access to longitudinal
patient-level information. Unlike cross-sectional datasets, these longitudinal records preserve the temporal ordering of treatment, mediator, and outcome required for causal mediation analysis, providing a foundation for mechanism-oriented research.

However, mediation analyses conducted within a single medical center are often underpowered, particularly for rare outcomes, specific treatment subgroups, or pathways requiring sufficient variability in both mediator and outcome measurements. Multi-institutional data can substantially improve statistical power and enable more generalizable mechanism-focused analyses. However, centralized pooling of patient-level data is  often constrained by privacy regulations, institutional governance policies, data-use agreements, and technical barriers to harmonized data transfer \citep{froelicher2021,vogelsang2020,bonomi2020}. These challenges have motivated federated learning approaches, which enable sites to collaborate by sharing low-dimensional aggregate information rather than individual-level records, while preserving local control over sensitive data \citep{tong2022,duan2020odal,spath2022,luo2024online}.

Communication-efficient federated methods for distributed healthcare data have been developed and largely fall into two categories: surrogate likelihood approaches \citep{duan2019odal,duan2020odal,duan2020,luo2022odach} and sequential renewable estimation frameworks \citep{luo2020}. These approaches allow sites to share summary-level information for joint analysis and have been adapted to a range of statistical settings, including continuous, binary, and survival outcomes \citep{tong2022,luo2020,lu2015,li2023,lu2021,jang2025,malcolm2025}. All these methods have been shown to achieve statistical efficiency comparable to pooled analyses of individual-level data. To our knowledge, no existing method enables causal mediation analysis in a federated setting, leaving a critical methodological gap for investigating treatment mechanisms across distributed healthcare networks.


Built on the counterfactual mediation framework \citep{robins1992,pearl2001,vanderweele2015,vansteelandt2012}, we develop a privacy-preserving approach for estimating causal mediation effects across distributed healthcare networks. This approach consists of three steps. First, a global outcome model is estimated across sites using renewable estimating equations \citep{luo2020}, and the fitted model coefficients are returned to participating sites. Second, each site locally constructs an expanded dataset by duplicating each observation, assigning the observed treatment in one copy and the alternative treatment in the other, and using the fitted outcome model obtained from Step 1 to impute the corresponding counterfactual outcomes. Finally, a global natural effect model is estimated from the expanded local datasets using renewable estimating equations. Throughout the process, sites exchange only low-dimensional aggregate quantities required for the renewable updates, such as updated parameter estimates and cumulative Hessian summaries. Because the expanded-data construction and counterfactual outcome generation are performed entirely within each institution, neither patient-level records nor expanded observations are shared. 

We applied the proposed federated mediation framework to investigate the extent to which the glycemic benefit of glucagon-like peptide-1 receptor agonists (GLP-1 RAs) is mediated through change in body mass index (BMI), used as a clinically available measure of weight-related change, among patients with type 2 diabetes. GLP-1 RAs lower glycated hemoglobin (HbA1c) through multiple pancreatic and extrapancreatic mechanisms. These include glucose-dependent insulin secretion, glucagon suppression, delayed gastric emptying, reduced appetite and food intake, and downstream metabolic effects associated with weight reduction \citep{muller2019glucagon,cornell2020review}. Although GLP-1 RAs reduce both HbA1c and body weight \citep{yeh2023effect,yao2024comparative}, these responses are not always proportional. This suggests that glycemic improvement may occur through both weight-mediated and non-weight-mediated pathways.

Quantifying the BMI-mediated component of HbA1c reduction helps distinguish glycemic benefit associated with weight-related change from benefit attributable to other mechanisms. This distinction is particularly relevant in routine clinical practice, where patients differ in baseline BMI, diabetes severity, comorbidities, medication adherence, concomitant therapies, and follow-up patterns. Prior mediation analyses of tirzepatide, a dual GIP/GLP-1 receptor agonist, have shown that HbA1c reductions are explained by both weight-loss-dependent and weight-loss-independent pathways, with the estimated contribution of weight loss varying across trial populations and comparators \citep{vilsboll2025hba1c}. However, these findings were derived from randomized clinical trial cohorts with controlled eligibility criteria and may not fully reflect treatment responses in routine clinical practice.

In this study, we use large-scale real-world data from the Indiana Network for Patient Care (INPC), a regional health information exchange that integrates longitudinal clinical information from multiple health systems across Indiana, including electronic health records, laboratory results, medication records, diagnoses, and encounter data. The breadth and longitudinal structure of INPC data make it well-suited for evaluating GLP-1 RA treatment effects in a heterogeneous population receiving care in routine clinical settings \citep{williams2025evolution}. 

We treat change in BMI as the mediator and decompose the total treatment effect on HbA1c into a natural indirect effect operating through BMI reduction and a natural direct effect representing all remaining pathways. This decomposition provides insight into the relative contributions of weight-mediated and weight-independent mechanisms. Furthermore, because INPC supports both distributed and centralized data access, it provides an ideal setting for validating the proposed federated framework, with success defined by close agreement between federated and pooled-data analyses.

\section{Results}
\label{sec:results}

\subsection{Evaluate federated mediation analysis approach by simulations}
\label{sec:sim_results}

We evaluated the proposed federated mediation estimator in Monte Carlo simulations. Distributed networks were generated with $K$ (5, 10, or 20) participating sites, and the number of patients at each site was drawn independently from a uniform distribution between 50 and 1,000. For each patient, we generated ten baseline confounders, a binary treatment indicator, repeatedly measured continuous mediator and outcome variables.

\begin{figure}[!ht]
\centering
\includegraphics[width=\textwidth]{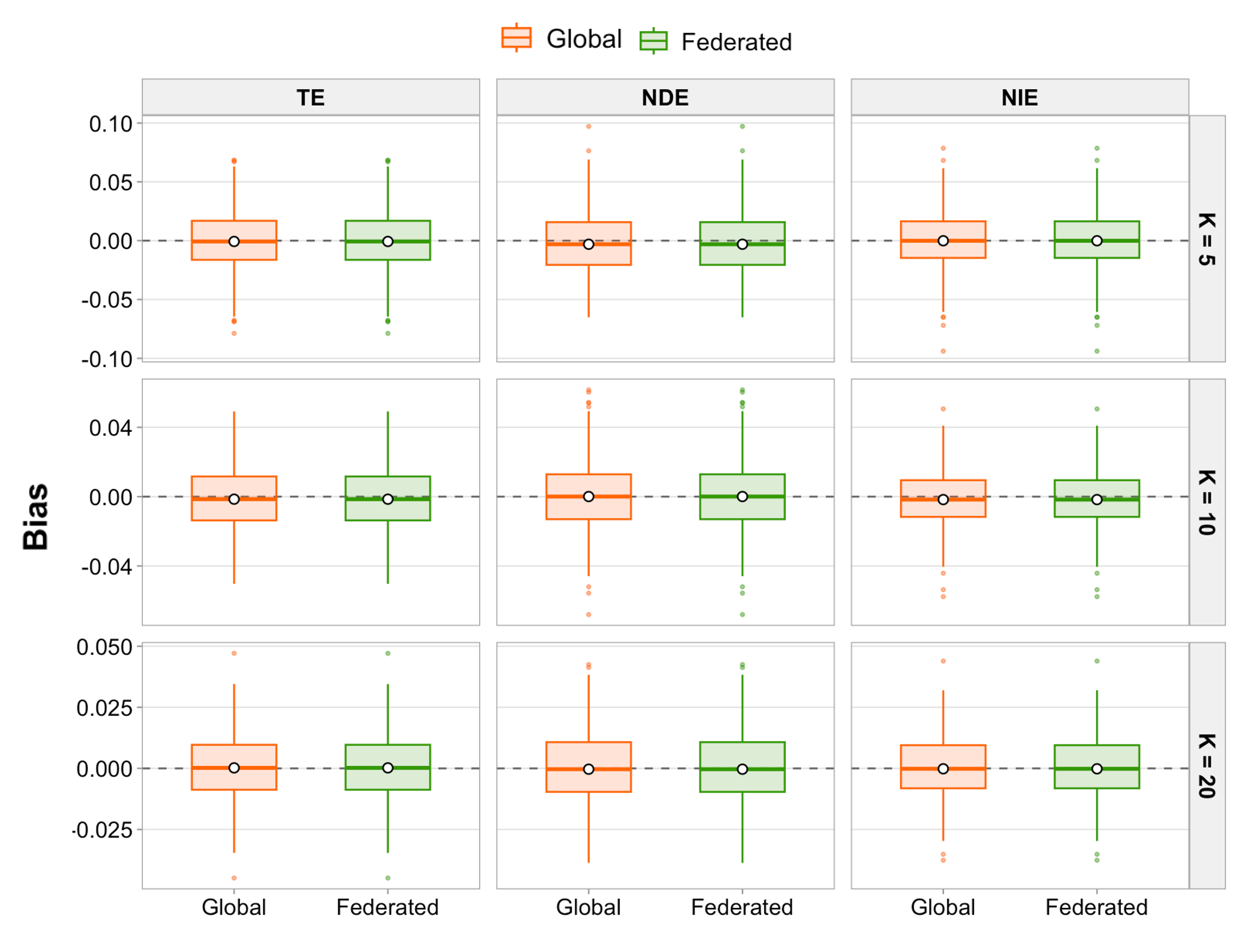}
\caption{\textbf{Empirical bias across 500 Monte Carlo replications.} Boxplots of bias (estimate $-$ truth) for the global (pooled, orange) and federated (FL, green) estimators, stratified by number of sites ($K \in \{5, 10, 20\}$, rows) and estimand (TE, NDE, NIE, columns). The dashed horizontal line indicates zero.}
\label{fig:bias_boxplot}
\end{figure}

\begin{figure}[!ht]
\centering
\includegraphics[width=\textwidth]{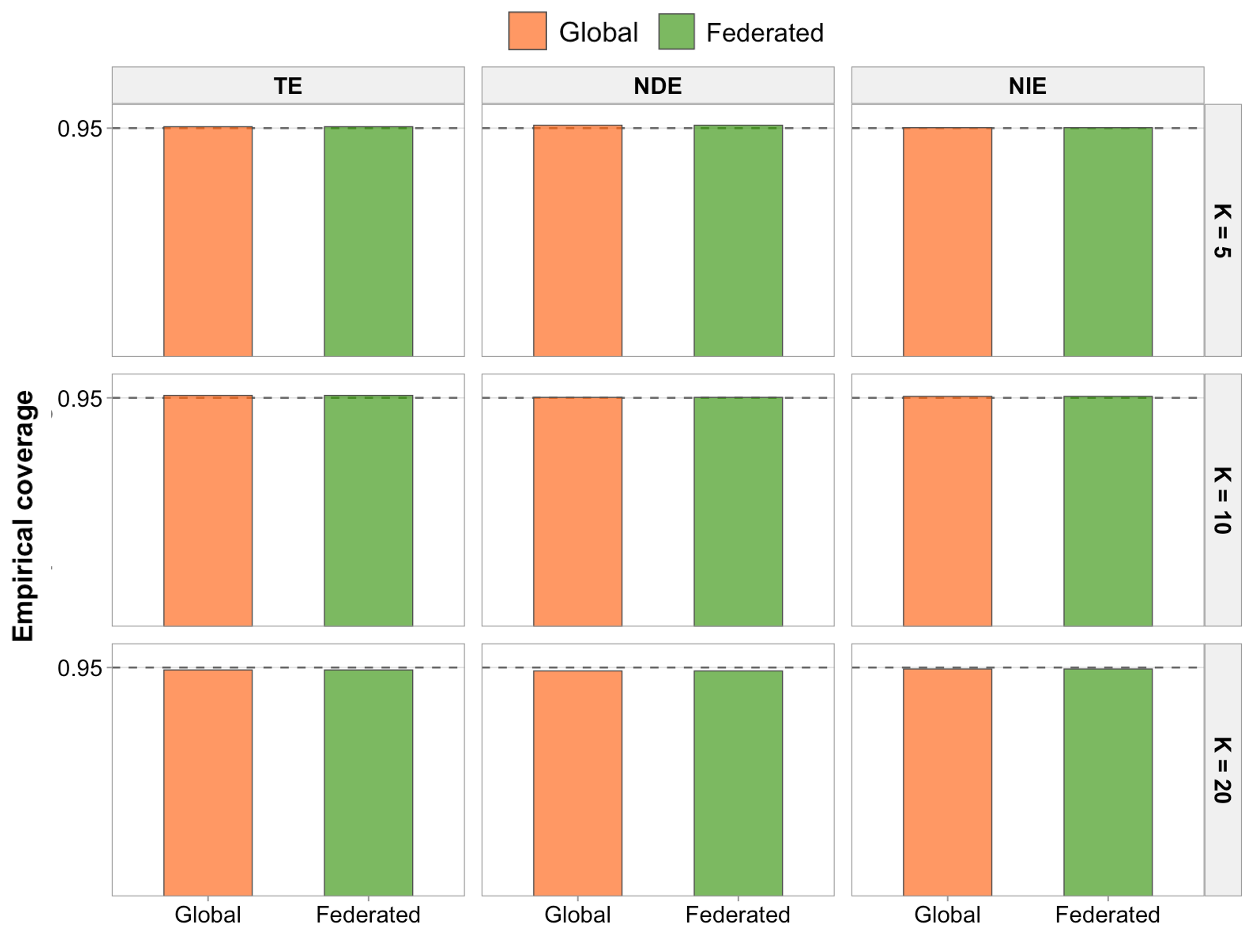}
\caption{\textbf{Empirical 95\% confidence interval coverage.} Bar charts of empirical coverage for the global (pooled, orange) and federated (FL, green) estimators, stratified by number of sites ($K \in \{5, 10, 20\}$, rows) and estimand (TE, NDE, NIE, columns). The dashed horizontal line indicates the nominal 0.95 level.}
\label{fig:coverage}
\end{figure}

Across all network configurations, the proposed federated mediation method produced unbiased estimates of the natural direct effect (NDE), natural indirect effect (NIE), and total effect (TE) (Figure~\ref{fig:bias_boxplot}). Its mean estimates closely matched the pooled benchmark analysis, without requiring patient-level data sharing. The 95\% confidence interval coverage remained close to the nominal level across all values of $K$, ranging from 0.94 to 0.96 (Figure~\ref{fig:coverage}). These findings support that, under the simulated settings, the proposed approach can closely reproduce pooled-data performance with valid uncertainty quantification, while avoiding patient-level data sharing.

\subsection{Structural challenges of cross-site patient overlap in federated networks}
\label{sec:overlap_simulation}

In privacy-preserving distributed EHR analyses, participating institutions do not share protected identifiers or patient-linkage information. As a result, when the same individual receives care at multiple sites, records from different institutions cannot be directly linked and are typically treated as independent observations. This challenge is not unique to our method but is a fundamental limitation of federated analyses based on distributed EHR data.

To evaluate the impact of cross-site patient overlap on causal mediation analysis, we conducted additional simulation studies with varying levels of overlap across sites. The overlap proportion was calculated as the percentage of unique patients appearing in two or more site groups. In the simulation study, this proportion was set to 10\%, 20\%, or 30\% of the total unique patient population. We compared three analytical strategies: (1) a pooled analysis that treats records from the same patient at different sites as independent observations, reflecting the assumption inherent in federated analyses; (2) the proposed federated mediation estimator, which does not require cross-site patient linkage and therefore operates on the same data structure as (1); and (3) a pooled analysis that correctly identifies and accounts for duplicate records from the same patient across sites, serving as the centralized reference benchmark.

The three analytical strategies produced unbiased mean estimates for the TE, NDE, and NIE across all levels of cross-site patient overlap. The main impact of unlinked cross-site records was on uncertainty quantification. Compared with the pooled analysis with cross-site records linked, the pooled and federated analyses with cross-site records unlinked produced narrower confidence intervals and lower empirical coverage, with under-coverage becoming more apparent as patient overlap increased from 10\% to 30\%. Because cross-site records from the same patient induce within-patient correlation, treating them as independent leads to underestimated standard errors, which in turn produces narrower confidence intervals and reduced coverage. This pattern indicates that unresolved cross-site patient overlap primarily affects variance estimation rather than mean estimation in privacy-preserving distributed EHR analyses.

\begin{figure}[!ht]
\centering
\includegraphics[width=1\textwidth]{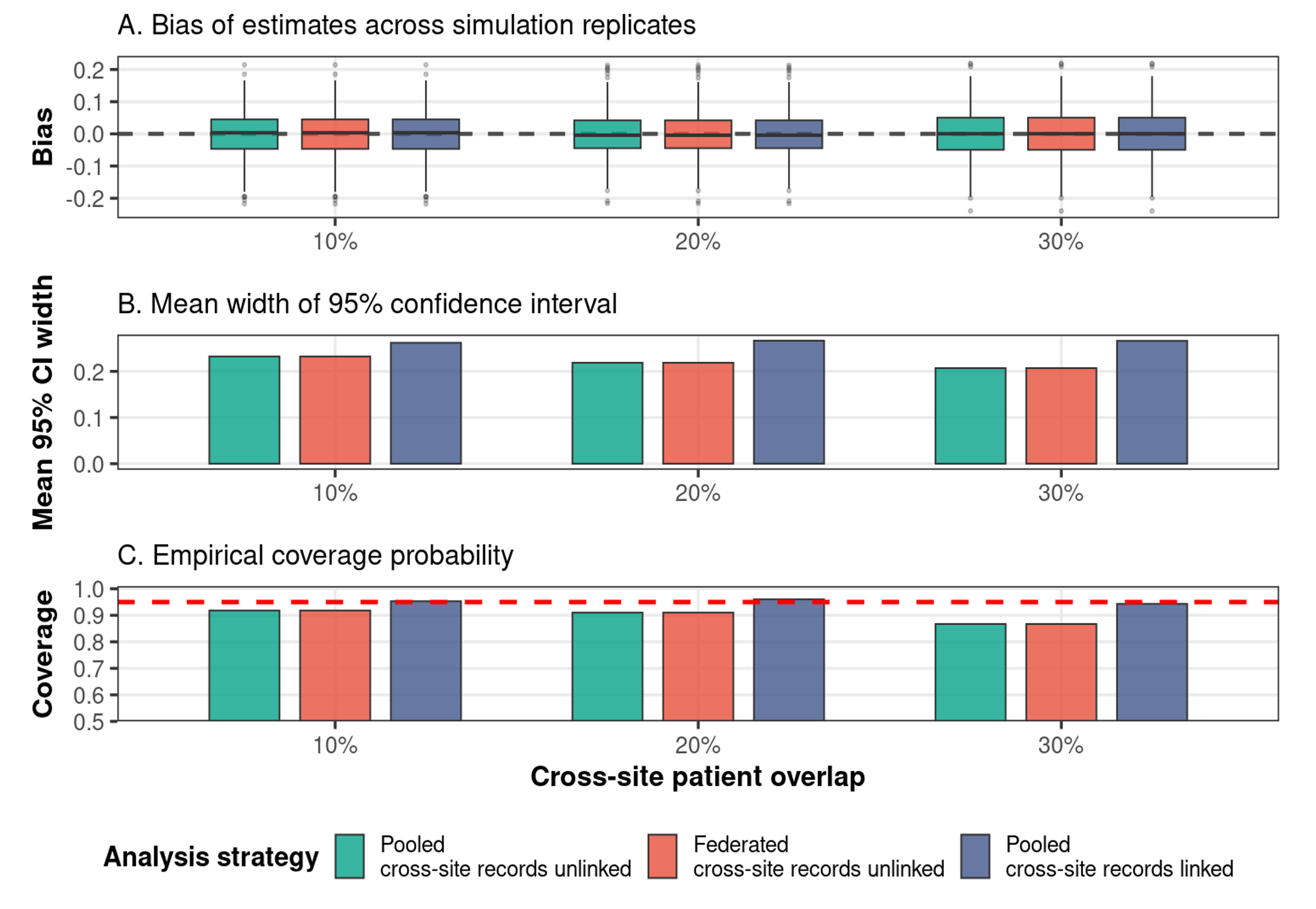}
\caption{\textbf{Impact of cross-site patient overlap on estimation of the natural indirect effect.} Simulation results for the natural indirect effect (NIE) across increasing levels of cross-site patient overlap (10\%, 20\%, and 30\%). Panel A shows bias across simulation replicates, with the horizontal dashed line indicating zero bias. Panel B shows the mean width of the 95\% confidence interval. Panel C shows empirical 95\% confidence interval coverage, with the horizontal dashed line indicating the nominal 0.95 level. The pooled analysis with cross-site records linked serves as the centralized benchmark, in which records from the same patient across sites are identified as belonging to the same individual and accounted for in variance estimation. The pooled and federated analyses with cross-site records unlinked treat records observed at different sites as independent.}
\label{fig:overlap_nie}
\end{figure}

Figure~\ref{fig:overlap_nie} illustrates this pattern for the NIE; the corresponding TE and NDE results exhibited the same qualitative pattern. The results highlight an important limitation of privacy-preserving distributed EHR analyses: when cross-site linkage is unavailable, patient mobility across institutions may lead to underestimation of variance and under-covered confidence intervals.

\subsection{Evaluate federated mediation analysis of GLP-1 RAs on HbA1c by INPC data}
\label{sec:real_data}

We applied the proposed federated mediation approach to EHR data from the Indiana Network for Patient Care (INPC), comprising records from 10 regional hospitals. The cohort included 32,146 patients with type 2 diabetes, of whom 8,687 initiated a glucagon-like peptide-1 receptor agonist (GLP-1 RA) and 23,459 initiated other glucose-lowering drugs. We considered change in body mass index ($\Delta \mathrm{BMI}$, kg/m$^2$) as the mediator and change in glycated hemoglobin from baseline ($\Delta \mathrm{HbA1c}$, percentage points) as the clinical outcome.

Baseline characteristics are summarized in Table~\ref{tab:table1}.  Compared with initiators of other glucose-lowering drugs, GLP-1 RA initiators had higher baseline BMI (mean 35.84 vs. 33.28 kg/m$^2$) and higher baseline HbA1c (mean 8.28\% vs. 7.88\%). We adjusted all 27 baseline variables listed in Table~\ref{tab:table1} to reduce bias due to systematic differences between treatment groups. These variables were treated as potential confounders because they may influence both the likelihood of GLP-1 RA initiation and subsequent changes in HbA1c and BMI. For example, clinicians may be more likely to prescribe GLP-1 RAs to patients with higher baseline BMI or HbA1c, or to those with cardiovascular or kidney disease \citep{american20269}. At the same time, these characteristics are also associated with differences in glycemic trajectories and weight change. Without adjustment, the estimated treatment effect could therefore be biased by confounding by indication.

\begin{table}[!ht]
\centering
\caption{\textbf{Baseline characteristics of the study cohort.} Data represent the 32,146 distinct patients who received care at 10 hospitals in the INPC network, stratified by GLP-1 RA initiation.}
\label{tab:table1}
\resizebox{0.95\textwidth}{!}{%
\begin{tabular}{lrrr}
\toprule

\textbf{Characteristic} & \textbf{GLP-1 RA initiators ($N=8,687$)} & \textbf{Other glucose-lowering drug initiators ($N=23,459$)} & \textbf{Total ($N=32,146$)} \\
\midrule
\multicolumn{4}{l}{\textit{\textbf{Demographics}}} \\
Age (years), mean (SD) & 61.84 (7.66) & 65.67 (9.30) & 64.63 (9.05) \\
Female & 4,982 (57.4\%) & 11,935 (50.9\%) & 16,917 (52.6\%) \\
Race/Ethnicity & & & \\
\quad Non-Hispanic White & 5,785 (66.6\%) & 14,599 (62.2\%) & 20,384 (63.4\%) \\
\quad Non-Hispanic Black & 1,089 (12.5\%) & 2,738 (11.7\%) & 3,827 (11.9\%) \\
\quad Hispanics & 537 (6.2\%) & 2,163 (9.2\%) & 2,700 (8.4\%) \\
\quad Others & 1,276 (14.7\%) & 3,959 (16.9\%) & 5,235 (16.3\%) \\
\midrule
\multicolumn{4}{l}{\textit{\textbf{Clinical Measurements}}} \\
Baseline HbA1c (\%), mean (SD) & 8.28 (1.79) & 7.88 (1.67) & 7.99 (1.72) \\
Baseline BMI (kg/m$^2$), mean (SD) & 35.84 (6.80) & 33.28 (5.96) & 33.97 (6.30) \\
\midrule
\multicolumn{4}{l}{\textit{\textbf{Comorbidities \& Medical History}}} \\
Hypertension & 7,122 (82.0\%) & 19,086 (81.4\%) & 26,208 (81.5\%) \\
Hyperlipidemia & 6,546 (75.4\%) & 17,076 (72.8\%) & 23,622 (73.5\%) \\
Cardiovascular Disease & 2,633 (30.3\%) & 6,958 (29.7\%) & 9,591 (29.8\%) \\
Thyroid Disease & 2,199 (25.3\%) & 5,070 (21.6\%) & 7,269 (22.6\%) \\
Depression & 2,139 (24.6\%) & 3,916 (16.7\%) & 6,055 (18.8\%) \\
Diabetic Neuropathy & 1,954 (22.5\%) & 3,530 (15.0\%) & 5,484 (17.1\%) \\
Chronic Kidney Disease & 1,500 (17.3\%) & 4,090 (17.4\%) & 5,590 (17.4\%) \\
Heart Failure & 1,112 (12.8\%) & 2,603 (11.1\%) & 3,715 (11.6\%) \\
Asthma & 1,092 (12.6\%) & 2,086 (8.9\%) & 3,178 (9.9\%) \\
Peripheral Vascular Disease & 1,015 (11.7\%) & 2,739 (11.7\%) & 3,754 (11.7\%) \\
Cerebrovascular Disease & 869 (10.0\%) & 2,421 (10.3\%) & 3,290 (10.2\%) \\
Non-Alcoholic Fatty Liver Disease & 723 (8.3\%) & 1,225 (5.2\%) & 1,948 (6.1\%) \\
Atrial Fibrillation & 711 (8.2\%) & 2,174 (9.3\%) & 2,885 (9.0\%) \\
Hypoglycemia & 427 (4.9\%) & 816 (3.5\%) & 1,243 (3.9\%) \\
Diabetic Retinopathy & 392 (4.5\%) & 637 (2.7\%) & 1,029 (3.2\%) \\
Hyperglycemic Emergency & 280 (3.2\%) & 509 (2.2\%) & 789 (2.5\%) \\
Alcohol Use Disorder & 186 (2.1\%) & 537 (2.3\%) & 723 (2.2\%) \\
Mild Cognitive Impairment & 49 (0.6\%) & 166 (0.7\%) & 215 (0.7\%) \\
\midrule
\multicolumn{4}{l}{\textit{\textbf{Medication Use \& Behaviors}}} \\
Insulin & 3,692 (42.5\%) & 4,260 (18.2\%) & 7,952 (24.7\%) \\
Oral Steroids & 1,968 (22.7\%) & 4,820 (20.5\%) & 6,788 (21.1\%) \\
Baseline SGLT2i Use & 318 (3.7\%) & 315 (1.3\%) & 633 (2.0\%) \\
Ever Smoker & 539 (6.2\%) & 739 (3.2\%) & 1,278 (4.0\%) \\
\bottomrule
\end{tabular}%
}
\end{table}

The INPC analysis also highlighted an important feature of multi-institutional EHR networks: patients may receive care at more than one participating institution. Among the 32,146 distinct patients in the Criterion 1 cohort, 20,953 patients (65.2\%) received care at one hospital site, 8,322 (25.9\%) at two sites, 2,335 (7.3\%) at three sites, and 536 (1.7\%) at four or more sites. Overall, 11,193 patients (34.8\%) had records from more than one hospital site. This structure is directly relevant to federated analyses, where patient linkage across institutions is often unavailable or restricted.

To leverage longitudinal EHR data to improve statistical efficiency, we constructed repeated mediator-outcome pairs by linking each HbA1c measurement to the most recent preceding BMI measurement. Because BMI and HbA1c were measured at irregular intervals during routine care, we adjusted for both the HbA1c measurement time and the time lag between BMI and HbA1c assessments. Within-patient correlation arising from repeated mediator–outcome pairs was accommodated using the proposed federated sandwich variance estimator.

As in typical federated analyses of distributed EHR data, records from the same patient appearing at different hospitals were treated as independent observations. Under this cross-site records unlinked setting, the federated estimator yielded a total effect of $\widehat{\mathrm{TE}}_{\mathrm{FL}} = -0.29$ (SE: $0.031$, 95\% CI: $-0.353$, $-0.233$). The estimated natural direct effect was $\widehat{\mathrm{NDE}}_{\mathrm{FL}} = -0.282$ (SE: $0.031$, 95\% CI: $-0.342$, $-0.222$), whereas the BMI-mediated natural indirect effect was smaller in magnitude, $\widehat{\mathrm{NIE}}_{\mathrm{FL}} = -0.011$ (SE: $0.003$, 95\% CI: $-0.016$, $-0.005$). These estimates closely matched the corresponding pooled analysis results. The estimated proportion of the total effect mediated through BMI change, defined as $\widehat{NIE}_{FL}/\widehat{TE}_{FL}$, was 3.7\% (95\% CI: 1.7\%, 5.6\%). This suggests that BMI change accounted for a modest component of the HbA1c reduction associated with GLP-1 RA initiation compared with initiation of other glucose-lowering drugs. 
 
\begin{figure}[!ht]
\centering
\includegraphics[width=\textwidth]{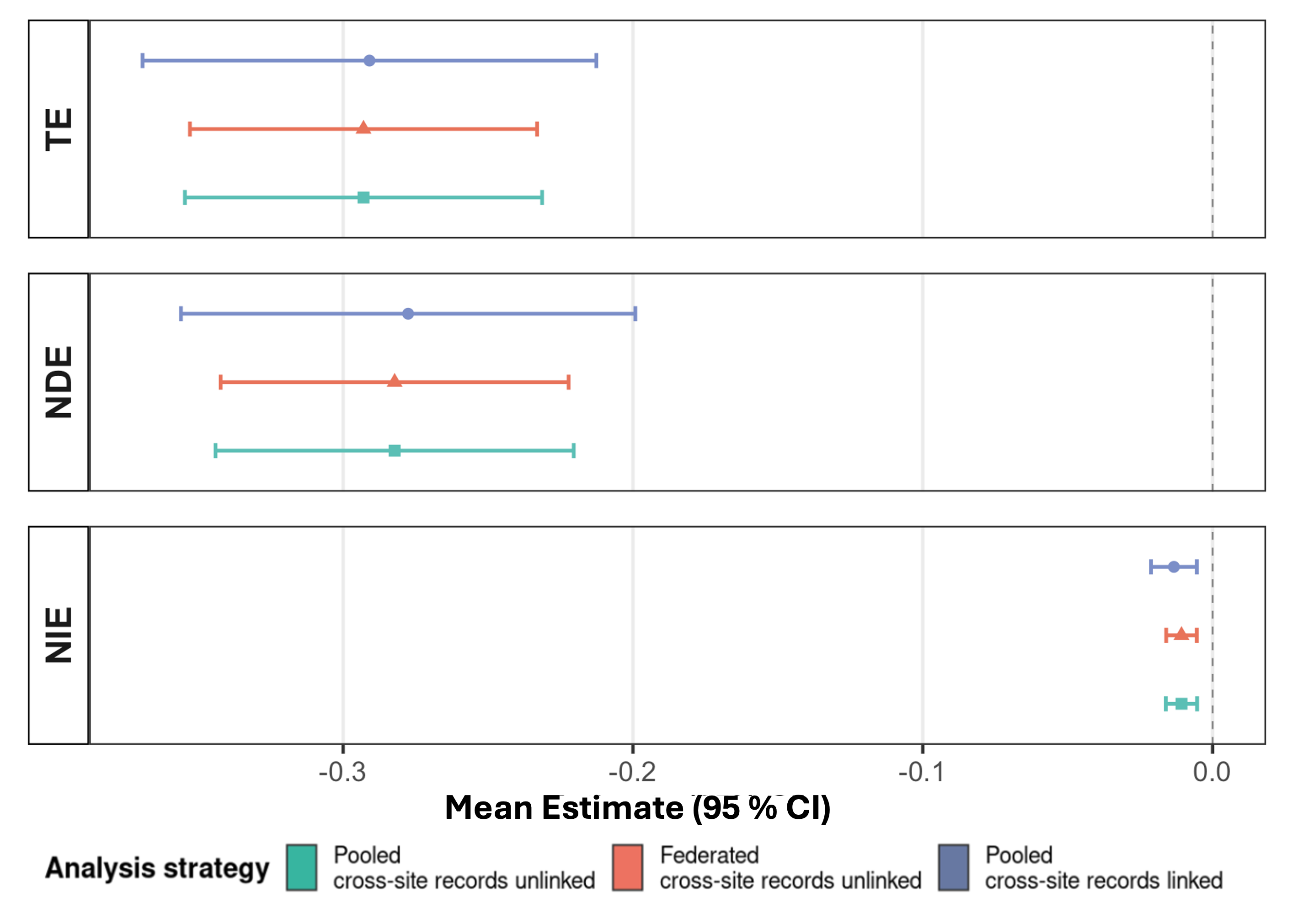}
\caption{\textbf{Real-data analysis by three analytical strategies.} Estimated total effect (TE), natural direct effect (NDE), and
natural indirect effect (NIE) of GLP-1 RA initiation on $\Delta$HbA1c, with $\Delta$BMI
as the mediator, from data across ten INPC hospitals. Points indicate mean estimates and
horizontal bars indicate 95\% confidence intervals. Squares: pooled analysis with cross-site records unlinked. Triangles: proposed federated estimator with cross-site records unlinked. Circles: pooled analysis with cross-site records linked to account for within-patient clustering across hospitals.}
\label{fig:realdata_forest}
\end{figure}

We also compared the federated results with two pooled analyses, as described in the simulation studies. The first pooled analysis used the same data structure as the federated analysis, treating cross-site records from the same patient as independent observations. The second pooled analysis leveraged patient linkage information to identify records belonging to the same individual across hospitals. Figure~\ref{fig:realdata_forest} presents estimates of the TE, NDE, and NIE from the three analytical strategies. Mean estimates were nearly identical across all three strategies, demonstrating that whether cross-site records were linked or unlinked did not bias pathway-specific mean estimates in this application. The primary difference was in uncertainty quantification: the federated estimator and the pooled analysis with cross-site records unlinked produced comparable standard errors, while linking records across hospitals in the pooled analysis yielded wider confidence intervals across all three estimands. For example, the standard error for the TE was $\widehat{\mathrm{SE}}_{\mathrm{FL}} = 0.031$ under the federated estimator and $\widehat{\mathrm{SE}}{\mathrm{pool}} = 0.040$ under the pooled analysis with cross-site records linked, with similar inflation observed for the NDE and NIE. These findings suggest that treating cross-site records from the same patient as independent observations may underestimate variance without distorting point estimates of the direct and indirect effects.

\section{Discussion}
\label{sec:discussion}

We introduced a privacy-preserving federated mediation method for estimating natural direct and indirect effects across distributed healthcare networks without sharing patient-level data. 
To our knowledge, this is the first federated method for causal mediation analysis, enabling natural direct and indirect effects to be estimated across distributed healthcare networks without transferring patient-level records.

The proposed framework addresses three core methodological challenges in extending 
causal mediation analysis to distributed EHR networks. First, we extend the three-stage natural effect estimation procedure to 
distributed data settings using renewable estimating equations \citep{luo2020}, so that each site transmits only updated parameter estimates and cumulative Hessian summaries required for renewable updating, without sharing any patient-level records. Second, we develop a robust sandwich variance estimator based on stacked estimating equations for the outcome and natural effect models, allowing uncertainty from the first-stage outcome model to be propagated into inference for the natural direct and indirect effects. Third, the method accommodates within-patient longitudinal dependence by locally aggregating patient-level score contributions within each site before constructing aggregate sandwich variance components, enabling valid and robust variance estimation for repeated mediator and outcome measurements.

Simulation studies showed that the proposed federated estimator achieved performance virtually identical to the pooled estimator across distributed networks with varying numbers of participating sites, producing unbiased estimates, accurate  variance estimates, and empirical coverage probabilities close to the nominal 95\% level for the natural direct, natural indirect, and total effects. The real-world application to the INPC data further demonstrated that, under the same patient-identification assumptions, the federated and pooled analyses yielded nearly identical mean estimates and confidence intervals. Together, these findings suggest that the proposed method can recover pooled-analysis results while preserving patient privacy in distributed healthcare networks.

We also investigated cross-site patient overlap, a common feature of distributed healthcare networks. In privacy-preserving federated analyses, patient identifiers are typically unavailable across institutions, preventing linkage of records belonging to the same individual. Consequently, when a patient receives care at multiple sites, their records are commonly treated as independent observations. Both simulation studies and the INPC application demonstrated that cross-site patient overlap has a negligible impact on mean estimation of natural direct and indirect effects; however, when cross-site dependence is ignored and overlapping records are treated as independent, the variance is systematically underestimated, resulting in confidence intervals that are narrower than they should be. The magnitude of this coverage gap increased with the degree of patient overlap, consistent with the growing covariance that is left unaccounted for as more records from the same individual are treated independently.

In the proposed implementation, we estimated the natural effect model using an outcome-regression-based natural effect imputation approach. As an alternative, inverse-probability-weighted (IPW) natural effect models\citep{vansteelandt2012} can be used to construct counterfactual contrasts by weighting observations according to estimated treatment and mediator-related mechanisms. We applied the IPW approach to simulated analyses; however, the resulting estimates exhibited greater variability due to extreme weights arising from imbalanced treatment allocation and large number of confounders. Consequently, compared with weighting-based 
methods --- which are susceptible to numerical instability from extreme 
propensity score weights and practical positivity violations --- estimating 
natural effect models (NEMs) via an outcome regression framework provides 
greater statistical efficiency and more robust inference 
\citep{vansteelandt2012,tchetgen2012}.

Our federated mediation approach has several limitations. First, the renewable estimating equations framework assumes that model parameters --- including the outcome regression coefficients and the natural 
effect model coefficients --- are homogeneous across participating sites. 
Site-level heterogeneity in  
treatment effect magnitude or mediator--outcome relationships across institutions cannot be accommodated under the current framework. Extending the renewable estimation procedure to 
incorporate site-specific random effects or stratified estimation 
strategies would broaden the applicability of the approach to more heterogeneous network settings. 

Second, the consistency of the federated estimator relies on correct 
specification of both the outcome regression model and the natural effect 
model at each site; misspecification of either component can induce 
persistent bias in the estimated natural direct and indirect effects that 
does not diminish with increasing sample size. Developing semiparametric 
or doubly robust extensions \citep{tchetgen2012,vansteelandt2012} --- 
which require correct specification of only one of the two component 
models --- would substantially strengthen the inferential robustness of 
the federated mediation framework, particularly in settings where the 
functional form of the outcome or mediator model is uncertain.

Finally, the current work focuses on a single mediator and a continuous outcome, which represents a practical scope given the complexity of the federated estimation procedure. The underlying counterfactual framework, however, can accommodate multiple mediators, diverse outcome types, and treatment–mediator interactions. Future extensions of the proposed federated framework will address jointly evaluated multiple correlated mediators, time-varying treatment strategies, and mediation analysis for survival and competing-risk outcomes—settings increasingly relevant to real-world EHR-based studies.

\section{Methods}

\subsection{Mediation Analysis at a Single Site}
We first define the causal estimands for mediation analysis. Let $A \in \{0,1\}$ denote treatment initiation, $M$ a post-treatment mediator, $Y$ the outcome, and $C$ a vector of baseline covariates. The nested counterfactual $Y(a,M(a^*))$ denotes the outcome that would be observed if treatment were set to $a$, while the mediator were set to the value it would naturally take under the observed treatment level $a^*$.

For a binary treatment, the total effect (TE) can be decomposed into the natural direct effect (NDE) and natural indirect effect (NIE) \citep{pearl2001,vanderweele2015,vansteelandt2012}:
\begin{equation}
\mathrm{TE} = E[Y(1)-Y(0)] = \underbrace{E[Y(1,M(0))-Y(0,M(0))]}_{\mathrm{NDE}}
+
\underbrace{E[Y(1,M(1))-Y(1,M(0))]}_{\mathrm{NIE}} .
\end{equation}

The NDE captures the component of the treatment effect not operating through the mediator, comparing treatment versus control while holding the mediator to the value it would naturally take under control. The NIE captures the component operating through the mediator, comparing the mediator values that would naturally arise under treatment versus control while holding treatment fixed at the treated level.

The estimation proceeds in three steps. First, we fit an outcome regression model for $E(Y \mid A,M,C)$ using the observed data. Second, we construct an expanded dataset for natural effect imputation by creating replicated observations for each individual. Each replicated row indexes a counterfactual treatment value $a$ while retaining the mediator value corresponding to the treatment level actually observed, denoted by $a^*=A_i$. For rows in which $a=A_i$, the observed outcome is retained. For rows in which $a\neq A_i$, the counterfactual outcome is imputed from the fitted outcome model as $ Y_i^\dagger = \widehat{E}(Y \mid A=1-A_i, M=M_i, C=C_i).$

Finally, we fit the natural effect model to the expanded dataset:
\begin{equation}\label{eq:nei_model}
    g\{E(Y^\dagger \mid a,a^*,C)\} = \eta_0 + \eta_a a + \eta_{a^*} a^* + \eta_C^\top C.
\end{equation}
With the identity link, $\eta_a$, $\eta_{a^*}$, and $\eta_a+\eta_{a^*}$ estimate the NDE, NIE, and TE, respectively.

\subsection{Federated mediation analysis}
\label{sec:fed_renewable}

\begin{figure}[!ht]
\centering
\includegraphics[width=\textwidth]{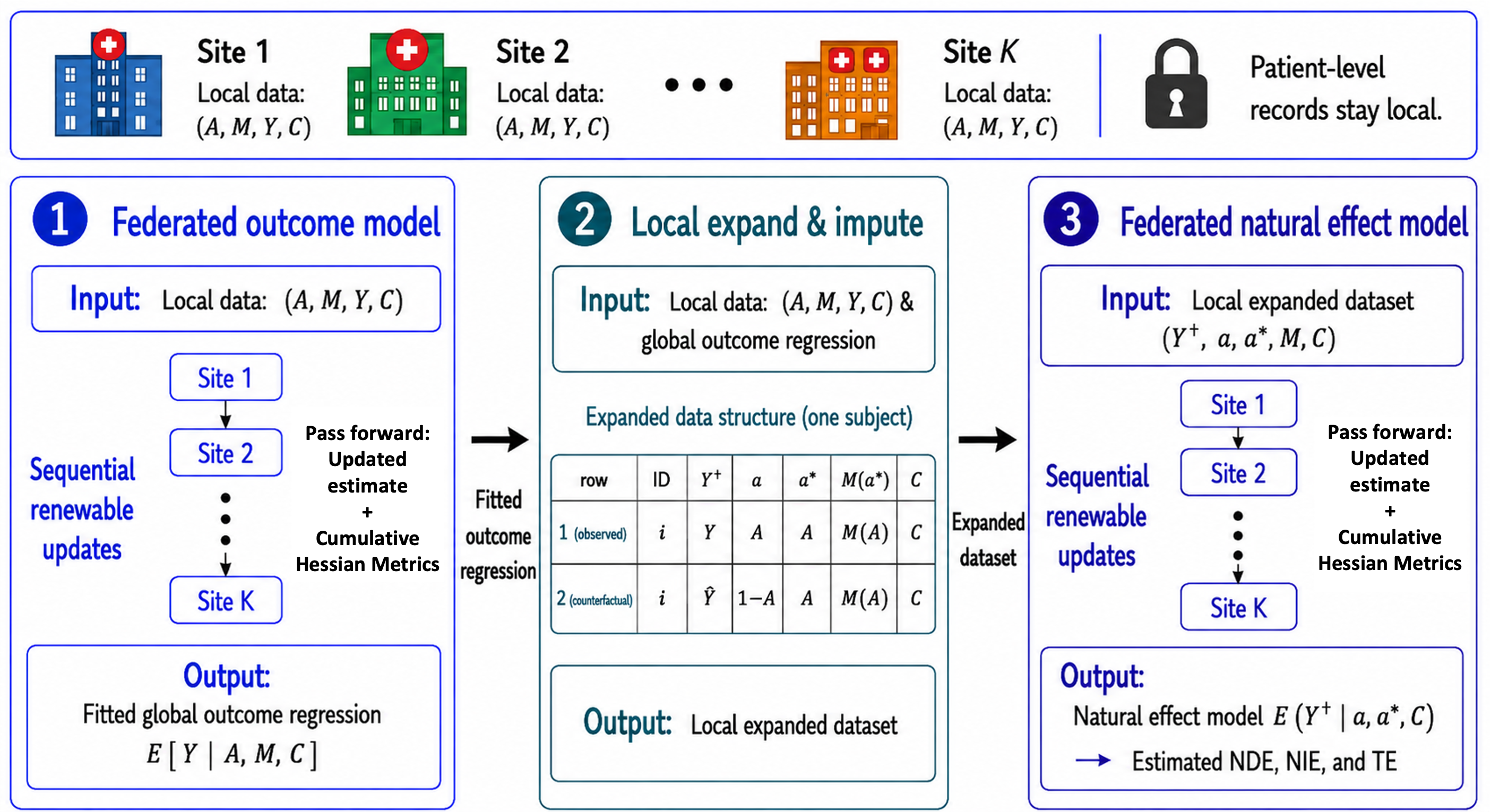}
\caption{\textbf{Overview of the federated mediation estimator.} The framework consists of three stages. First, a global outcome regression model is estimated across sites using sequential renewable updates, and the fitted model coefficients are broadcast to each site. Second, an expanded dataset is created at each local site. Third, the natural effect model is applied to the local expanded dataset using renewable updates. Patient-level EHR records remain local throughout the procedure, and the final outputs are estimates of the natural direct effect, natural indirect effect, and total effect.
\label{fig:workflow}}
\end{figure}

We propose a federated mediation framework that extends the counterfactual mediation approach described in Section 4.1 from a single site to distributed EHR networks.  The key idea is to apply sequential renewable estimating equations \citep{luo2020} to the two regression models above: first to the outcome model and then to the natural effect model. Patient-level records remain within each site throughout the procedure. As illustrated in Figure~\ref{fig:workflow}, the federated procedure follows the same three-step structure as the single-site estimator.

\begin{enumerate}
    \item \textbf{Federated outcome model fitting.}
    The outcome regression model, $E(Y\mid A,M,C)$, is fitted sequentially across sites using the renewable updating procedure. Each site receives the current coefficient estimate and cumulative Hessian summary from the previous site, computes the local score and Hessian using only local data to update the coefficient estimate, and passes only the updated coefficient estimate and updated cumulative Hessian summary to the next site. After the final update, the fitted global outcome regression model is broadcast back to all participating sites for subsequent local expansion and counterfactual imputation.

    \item \textbf{Local expansion and counterfactual imputation.} 
    Using the fitted global outcome model, each site constructs an expanded dataset by creating two rows for each patient as described in Section 4.1 and imputes the corresponding counterfactual outcomes locally. This expansion and imputation step is performed independently and in parallel across sites, and no expanded patient-level data are shared.

    \item \textbf{Federated natural effect model fitting.}
    The natural effect model is fitted by applying the same renewable updating procedure to the locally expanded and imputed datasets. The final global natural effect model yields estimates of NDE, NIE, and TE as described in Section 4.1.
\end{enumerate}

For statistical inference, we use a sandwich variance estimator to incorporate uncertainty arising from both the outcome regression and the subsequent natural effect model in \eqref{eq:nei_model}. Furthermore, we propose an efficient empirical sandwich variance estimator that accommodates longitudinal mediator and outcome measurements by aggregating score contributions at the patient level before constructing the sandwich estimator. In the federated implementation, each site computes its variance components locally, and only low-dimensional summaries are shared and combined to obtain the final variance estimate. This approach provides valid statistical inference while preserving patient privacy, as only updated parameter estimates, cumulative Hessian summaries, and aggregate variance components are exchanged across sites. Detailed estimating equations and variance derivations are provided in the Supplementary Methods.

\subsection{Simulation studies}

We conducted simulation studies to evaluate whether the proposed federated mediation method can reproduce the pooled-analysis benchmark, in which all individual-level data are analyzed centrally. The simulations generated distributed EHR data  from multiple sites, a binary treatment, baseline covariates, and longitudinally measured continuous mediators and outcomes.

For each simulated dataset, we applied our federated mediation method and compared it with the pooled-analysis benchmark. The pooled estimator served as the reference for assessing whether sequential renewable updates recovered the centralized natural effect analysis without transferring patient-level data. Performance was summarized using bias, empirical standard deviation, average estimated standard error, root mean squared error (RMSE), and empirical coverage of nominal 95\% confidence intervals.

We also evaluated cross-site patient overlap, a common feature of dense EHR networks in which the same person may receive care at multiple participating hospitals or clinics. In privacy-preserving federated analyses, patient identifiers are typically unavailable across sites, preventing linkage of records belonging to the same individual. As a result, records from the same patient at different institutions may be treated as independent observations. To assess the impact of this limitation, we conducted simulations with varying levels of patient overlap and examined its effects on both mean and variance estimation. Results were compared with those from an idealized pooled analysis that could correctly identify and account for all records belonging to the same individual across sites.

\subsection{Real-world data analysis using INPC}

We applied the proposed federated mediation analysis to EHR data from the Indiana Network for Patient Care (INPC), consisting of 10 regional hospitals. The study cohort included adults with type 2 diabetes who had complete data on treatment, BMI, HbA1c, and baseline covariate information. The exposure was GLP-1 RA initiation. The comparator was initiation of other glucose-lowering drugs.GLP-1 RAs included albiglutide, dulaglutide, exenatide, liraglutide, lixisenatide, semaglutide, and tirzepatide. Comparator glucose-lowering drugs included dipeptidyl peptidase-4 (DPP-4) inhibitors, sulfonylureas, thiazolidinediones, meglitinides, and alpha-glucosidase inhibitors.

The mediator was BMI change from baseline ($\Delta$BMI, kg/m$^2$), used as a weight-related pathway measure, and the outcome was HbA1c change from baseline ($\Delta$HbA1c, percentage points). Longitudinal BMI and HbA1c were measured during a one-year follow-up window. During the window, mediator-outcome pairs were constructed as follows: Working backward from the last available HbA1c measurement within the follow-up window, each HbA1c value was linked to the most recent preceding BMI measurement, resulting in one or more mediator-outcome pairs per patient. Constructing multiple pairs per patient made full use of the available longitudinal data and increased statistical power. To account for irregular observation schedules in routine clinical care, both the outcome measurement time and the lag between BMI and HbA1c assessments were included as covariates.

We adjusted for 27  confounders measured at baseline (Table~\ref{tab:table1}): age, sex, race/ethnicity, baseline BMI, baseline HbA1c, 18 comorbidities, 3 medication use indicators (insulin, oral steroids, and baseline SGLT2i use), and smoking status. Using the proposed federated mediation framework, we estimated the total effect (TE), natural direct effect (NDE), and natural indirect effect (NIE) of GLP-1 RA initiation on changes in HbA1c, with changes in BMI serving as the mediator. All effects were reported on the change scale. The NDE represents the mean difference in $\Delta$HbA1c associated with GLP-1 RA initiation not operating through BMI change, whereas the NIE represents the component of the effect operating through GLP-1 RA--associated changes in BMI.

The primary federated analysis treated records from the same patient observed at different hospitals as independent observations, reflecting a distributed setting in which cross-site patient identifiers are not shared. Within each site, repeated mediator--outcome observations from the same patient were accounted for using cluster-robust sandwich variance estimation.

We compared the federated estimates with two pooled analyses. The first pooled analysis used the same data structure as the federated analysis, treating records from the same patient across hospitals as independent observations. This comparison assessed whether the proposed federated mediation method could reproduce centralized natural effect estimation under the same working independence assumptions. The second pooled analysis leveraged patient linkage information to identify records belonging to the same individual across hospitals, allowing us to evaluate the impact of cross-site patient overlap on direct and indirect effect estimation. Mean estimates, standard errors, and two-sided 95\% confidence intervals were reported for all analyses. All analyses were implemented in R.

\FloatBarrier
\clearpage

\noindent\textbf{Data availability}

The data that support the findings of this study were obtained from the Regenstrief Institute Inc. via the Regenstrief Data Core, which provides access to data from the Indiana Network for Patient Care (INPC). Due to data use agreements, the data are not publicly available. However, interested researchers can request access to the data directly from the Regenstrief Institute Inc. for replication purposes.

\medskip

\noindent\textbf{Author contributions}

H.J. developed the federated mediation methodology, implemented the algorithms, conducted simulation studies, performed statistical analyses, interpreted the results, and drafted the manuscript. Y.L. contributed to statistical analyses and preparation of the real-world EHR application. R.R. contributed to interpretation of the clinical findings and manuscript writing. J.B. and S.G. provided data access and clinically interpreted the real-world EHR application. M.R., X.S., and L.Z. supervised the study, and provided methodological oversight. All authors read and approved the final manuscript.

\medskip

\noindent\textbf{Competing interests}

The author(s) declare no competing interests.

\medskip

\noindent\textbf{Ethics approval and consent to participate}

The study protocol was reviewed by the Institutional Review Board (IRB) of Indiana University (Protocol No. 27345). The IRB determined this study to be exempt from requiring oversight and waived the requirement for informed consent due to the exclusive use of de-identified data.

\medskip

\noindent\textbf{Funding}

Research reported in this publication was supported by the National Institute of Allergy and Infectious Diseases of the National Institutes of Health under Award Number R01AI158543. The content is solely the responsibility of the authors and does not necessarily represent the official views of the National Institutes of Health. 

\bibliographystyle{naturemag}
\bibliography{citations}
\end{document}